\documentclass[twocolumn,conference]{IEEEtran}
\usepackage[T1]{fontenc}
\usepackage[latin9]{inputenc}
\usepackage{float}
\usepackage{graphicx}

\makeatletter

\providecommand{\tabularnewline}{\\}

\usepackage[caption=false,font=footnotesize]{subfig}

\makeatother

\begin{document}
\title{Emergence of Roles in Robotic Teams with Model Sharing and Limited
Communication}
\author{\IEEEauthorblockN{Ian O'Flynn}\IEEEauthorblockA{EEE Department\\
Trinity College Dublin\\
Dublin, Ireland\\
Email: oflynni@tcd.ie}\and \IEEEauthorblockN{Harun Šiljak}\IEEEauthorblockA{EEE Department\\
Trinity College Dublin\\
Dublin, Ireland\\
Email: harun.siljak@tcd.ie}}
\maketitle
\begin{abstract}
We present a reinforcement learning strategy for use in multi-agent
foraging systems in which the learning is centralised to a single
agent and its model is periodically disseminated among the population
of non-learning agents. In a domain where multi-agent reinforcement
learning (MARL) is the common approach, this approach aims to significantly
reduce the computational and energy demands compared to approaches
such as MARL and centralised learning models. By developing high performing
foraging agents, these approaches can be translated into real-world
applications such as logistics, environmental monitoring, and autonomous
exploration. A reward function was incorporated into this approach
that promotes role development among agents, without explicit directives.
This led to the differentiation of behaviours among the agents. The
implicit encouragement of role differentiation allows for dynamic
actions in which agents can alter roles dependent on their interactions
with the environment without the need for explicit communication between
agents.
\end{abstract}

\IEEEpeerreviewmaketitle{}

\section{Introduction}

The use of multi-agent reinforcement learning (MARL) in practical
applications may face significant challenges primarily due to the
high energy consumption and computational demands in training models
{[}1{]}. The energy requirements of MARL systems, specifically those
utilising large deep learning models, can be prohibitive, limiting
their deployment in energy-sensitive environments such as mobile robotics
and embedded systems {[}2, 3{]}. Traditional reinforcement learning
(RL) algorithms show inefficiency in terms of computational demand
and power uses, particularly when scaled to larger systems. There
has also been a greater push in exploring energy efficient solutions
in MARL and other deep learning applications such as deep-neural networks
{[}4, 5{]}. By designing solutions to RL problems with higher levels
of energy efficiency, the use of RL in multi-agent environment systems
can be extended to a broader range of applications ensuring more sustainable
and versatile deployments. However, it should be ensured that the
shift towards energy efficient solutions does not compromise on the
learning and decision-making capabilities of the RL agents. 

Many multi-agent systems require agents to undertake different roles
within the environment, such as in RoboCup soccer, where robots are
assigned specific roles to maximise team performance {[}6{]}. This
role assignment may be achieved through inter-agent communication
and coordination {[}7{]}, or through implicit role development {[}8{]}.
These communication or design overheads can become substantial, and
act as a hindrance to the performance as the number of agents in the
environment increases. The reliance on the explicit definition and
management of roles among agents, through communication or definition,
creates a rigid structure in the population and can hamper adaptability
and scalability. For these approaches to successfully develop complex
behaviours and strategies the overhead of the solution significantly
increases {[}9{]}, but by implementing a reward function that allows
for implicit role definition, this increase can be mitigated while
promoting flexible and scalable agent behaviours.

In this paper, we propose a single-agent learning approach designed
to reduce both the energy and computational demands of MARL systems
in the case of resource foraging. This proposed solution leverages
a model sharing mechanism where a single agent\textquoteright s learned
strategies are disseminated across the population of non-learning
agents. This approach not only conserves energy by reducing the number
of agents that engage in the computationally intensive learning process
but also minimises the need for inter-agent communication by limiting
the real-time data exchange to essential model updates. By reducing
the number of learning agents in the environment, the time required
for training in this environment will be reduced. Additionally, we
introduce a reward function designed to facilitate implicit role development
among agents, removing the need for explicit role definitions or cooperative
communication and the associated overhead. The proposed reward function
rewards the agent based on their success in resource collection and
their proximity to other agents, which aims to encourage agents to
naturally assume roles that optimise the collective performance, relying
on environmental cues and agent interaction rather than predefined
assignments. While the effectiveness of using reward functions to
guide cooperative and competitive behaviors without fixed roles in
multi-agent settings has already been demonstrated {[}10{]}, this
approach extends the concept by integrating desired role specialisation
directly into the reward function.

\section{Background: learning to forage}

Q-Learning, a type of reinforcement learning, has advanced the field
of reinforcement learning by allowing agents to learn optimal actions
in a Markovian context without needing a model of the environment.
It has laid the foundation for further innovations such as Deep Q-Learning
(DQL), which integrates deep neural networks (DNNs) with Q-Learning
to handle more complex problems with high-dimensional state spaces.
These methods have not only propelled theoretical advancements but
also practical applications in robotics, gaming, and notably in solving
foraging problems {[}11{]}. 

Foraging problems, which involve searching for and gathering valuable
items or information in an environment, mirror many real-world challenges
across robotics and natural resource management {[}12{]}. The application
of Q-Learning and DQL to these problems has opened up new avenues
for research and optimisation in autonomous systems.

\section{System model}

\subsection{The environment}

The environment built for simulating and evaluating the performance
of reinforcement learning approaches to resource collection is created
using a square, discrete, grid-based spatial representation. The environment
is bounded by hard borders restricting the agent movement to the grid
boundaries. Hard borders were chosen over toroidal, or \textquoteright wrap
around\textquoteright{} borders, in which an agent attempting to move
off the edge of the grid would reappear on the corresponding cell
on the opposite edge, due to the computational simplicity and structure
they provide. The hard borders also better represent a real environment
than the torodial border, where the resources only appear in a given
area rather than infinite environment with no restrictions on movements.
All agents within the environment are limited to orthogonal movements
(up, down, left, right).

The environment contains a resource density parameter $\rho$, that
dictates the probability of a resource being present within any given
cell. The initial position of the ally and adversarial agents in the
environment at the beginning of every episode is determined using
a centred random distribution. The DQN agent is placed in the centre
of the environment for the initial episode and for subsequent episodes
follows the same distribution as the other agents.

To establish a performance baseline within the simulated environment,
adversarial agents are introduced into the environment to compete
with the reinforcement learning agents for the resources available
in the environment. These agents lack the capacity to learn, cooperate
or communicate so they provide a simplistic but comparative benchmark
to the proposed approach. They follow a simplified resource collection
algorithm that prioritises the closest visible resource. The choice
of a fixed, simple strategy helps us in maintaining a clear performance
baseline; in more complex scenarios, more advanced strategies for
the adversary team should be considered. In brief, we have two competing
teams of agents foraging a resource randomly distributed on a square
grid. In the remainder of the paper, ``agent'' usually refers to
a member of the team for which learning is happening, and not the
adversarial team.

\subsection{The learning}

The strategy chosen employs a single-agent reinforcement learning
strategy within the foraging environment containing adversarial agents
as an approach to solve a multi-agent reinforcement problem. Agents
are divided into \textquotedbl leaders\textquotedbl{} and \textquotedbl allies,\textquotedbl{}
with leaders engaging in continuous learning and policy refinement
through Deep Q-learning (DQL). Deep Q-Networks (DQNs) are chosen for
their proven ability to handle complex state spaces and process spatial
information, crucial for effective navigation and decision-making
in this context {[}11{]}. Leaders\textquoteright{} learned models
are periodically shared with allies, accelerating learning and promoting
successful strategies across the group. In this case a single leader
is chosen to learn in the environment. The use of a single learning
agent will see a large reduction in the training time of a population
and while aiming to match the performance of other approaches like
multi-agent reinforcement learning (MARL) and centralised networks.

The learning cycle of DQL is central to the leader agent\textquoteright s
ability to learn and adapt over time. It is a sequential process where
an agent continually refines its policy through direct interaction
with the environment. The cycle consists of observing the environment,
selecting actions, receiving rewards, and updating the internal Q-network
based on stored experiences. The input to the DQL process is a the
state of the environment perceived by the DQN agent. The environment
is represented as tensor with 3 different input layers, similar to
RGB channels of an image in typical convolutional neural network (CNN)
usage. The input layers in this case are state matrices representing
different aspects of the environment state. The locations of resources,
ally agents, and adversarial agents are each represented by a separate
state matrix. To process the state input perceived by the agent and
output the Q-values for each action in that state, a CNN, integrated
with fully connected layers, is used.

\subsection{The reward function}

The reward function employed for the single learning agent in the
environment rewards the agent for resource collection, proximity adversarial
agents, and distance from ally agents. The generalised reward function
is defined as: 

\[
R_{A_{i}}(a_{t})=R_{c}+R_{c}(w_{e}(R_{e}(A_{i}))+w_{a}(R_{a}(A_{i})))
\]
where: $R_{A}{}_{i}(a_{t})$ is the total reward for agent $A_{i}$
taking an action $a_{t}$, $R_{c}$ is the base reward for collecting
a resource, $R_{e}$ is the reward for proximity to adversarial agents,
$R_{a}$ the reward for distance from ally agents. $w_{e}$ weight
for the adversarial agents proximity reward, and $w_{a}$ weight for
the ally agent distance reward.

The reward for proximity to adversarial agents, $R_{e}$, and distance
from ally agents, $R_{a}$, are defined as 

\[
R_{e}(A_{i})=\sum_{j=1}^{n_{e}}1-\frac{d_{e}(A_{i},e_{j})}{D},\ R_{a}(A_{i})=\sum_{k=1}^{n_{a}}\frac{d_{a}(A_{i},a_{k})}{D}
\]
where: $n_{e}$ is the number of ally agents, $n_{a}$ number of adversarial
agents. $d_{e}(A_{i},e_{j})$ is the distance between agent $A_{i}$
and adversarial agent $e_{j}$ , $d_{a}(A_{i},a_{k})$ is the distance
between agent $A_{i}$ and ally agent $a_{k}$, and $D$ is the maximum
distance possible between an agent and an adversarial agent or ally
agent. Here, distance is taken in Manhattan distance sense.

These equations show the evaluation of the reward components $R_{e}$
and $R_{a}$. To promote agents collecting in areas in which ally
agents are not present, the additional reward is defined as the normalised
distance between the agent and all other ally agents. Therefore, as
the distance between the agent and its allies increases, the value
of $R_{a}$ also rises. This positive reinforcement for spatial separation
encourages resource collection away from other ally agents allowing
to mitigate competition for resources amongst allies.

The second additional reward element, $R_{e}$ , is defined as 1 minus
the normalised distance between the agent and all adversarial agents.
The inverse relationship between the distance and the value of $R_{e}$
means agents receive an increasing reward the closer they collect
a resource to an adversarial agent. This relationship between distance
and reward will encourage agents to operate in contestable ares, disrupting
the resource gathering efforts of the adversarial agents. The goal
of defining the total reward as the summation of $R_{c}$, $R_{e}$,
and $R_{a}$, was to foster strategic decision-making and the development
of specialised roles within the agent population. By balancing incentives
around proximity to adversarial agents and distance from ally agents,
the reward structure creates a complex optimisation landscape for
the agents. This landscape gives the opportunity for the development
of different agent roles within the environment. We can imagine a
possible scenario in which the roles of \textquoteright disruptors\textquoteright{}
and \textquoteright explorers\textquoteright{} are developed. The
ally agents and the leader can be carrying out two different actions
within the populations. The leader agent and two ally agents could
be operating in close proximity to the enemy agents, acting as disruptors
while two other ally agents are exploiting the resources away from
all other agents.

Implicit role development, as facilitated by the reward function,
offers several advantages over explicit role assignment in multi-agent
systems. The adaptability it allows is crucial in dynamic and unpredictable
environments, enabling agents to respond in real-time to changes and
optimise their strategies to current conditions rather than adhering
to potentially outdated predefined roles. This flexibility enhances
the system\textquoteright s overall resilience and effectiveness,
as agents continuously learn and evolve their behaviors to maximise
the collective reward. Scalability is another significant advantage
of implicit role definition. Managing and coordinating explicit roles
becomes increasingly complex and impractical as the number of agents
in the system grows. In contrast, implicit role development, driven
by individual reward maximization, naturally leads to role differentiation
without the need for centralised control or extensive coordination,
making the system more scalable and reducing the overhead associated
with role management.

Communication overhead is also significantly reduced in systems that
employ implicit role development. Explicit role assignment often requires
continuous communication among agents to coordinate their actions,
which can be burdensome, especially in constrained or distributed
environments. In contrast, implicit roles emerge from the agents\textquoteright{}
interactions and decisions based on localised information and the
reward structure, minimising the need for extensive communication.
Additionally, the learning and evolution aspect of implicit role development
means that agents are not static in their behaviors; they can improve
and refine their strategies over time. This continuous learning process,
driven by the reward function, allows for the optimisation of individual
and collective behaviors, leading to a more robust and efficient system.

\subsection{Model Sharing}

As stated when discussing the approach that this strategy takes, the
DQN agent, which serves as the primary learning entity, periodically
shares its learned model with the ally agents in the environment.
This sharing mechanism is designed to quickly disseminate successful
strategies across agents, elevating the collective performance. However,
to prevent uniformity in behavior and to foster a diverse range of
strategies among the ally agents, an evolutionary adaption process
is integrated into the model-sharing phase. The evolutionary adaption
introduces slight variations, or mutations, to the model parameters
during the sharing process. This leads to a population of ally agents
that, while benefiting from the learned experiences of the DQN agent,
also explores and develops unique strategies. This diversity is crucial
for the system\textquoteright s adaptability and resilience, allowing
the collective to tackle a broader spectrum of challenges. Model sharing
is executed at the end of every \textquoteright lifetime\textquoteright ,
defined as a series of $n$ episodes of learning and interaction within
the environment. During this process, each learnable parameter of
the DQN model is evaluated for potential mutation. For each parameter,
a noise tensor, matching the parameter\textquoteright s dimensions,
is generated. This noise represents the potential mutation and is
scaled by a hyperparameter. Consequently, ally agents not only learn
from the DQN agent but also develop their variations of the learned
behaviors. Hence, the system cultivates a range of behavioral types
among the agents, enhancing the collective\textquoteright s ability
to adapt to and thrive in the dynamic environment.

\section{Results and Discussion}

\subsection{Setup}

In order to determine the performance of the proposed strategy, the
DQN agent is put through a learning phase, which consists of $n_{L}$
lifetimes, each containing $n_{E}$ episodes. The model sharing occurs
between the DQN agent and the ally agents at the end of every lifetime.
There are no allies present in the first lifetime due to the lack
of a model being available to them so they are introduced at the beginning
of the second lifetime with the mutated version of the first learned
model. Each episode, $E$, consists of $n_{t}$ time steps. After
nt time steps the simulation environment is reset. The resources are
regenerated and all agent positions are reset. The performance statistics
of all agents are also reset to zero. Following the completion of
$n_{E}$ episodes, a lifetime has been completed. On completion of
the first lifetime, $n_{a}$ ally agents are added to the environment,
and the model learned in the initial lifetime is shared with some
mutated parameters. This process is repeated for the subsequent lifetimes,
until $n_{L}$ lifetimes have been completed. During each time step,
each leader and ally agent first gathers the information about the
environment within its visible area, including the locations of resources,
allies and adversaries. The leader agent will either explore the environment
with a random movement or exploit its learned model to select an action,
depending on the generated $\epsilon$ value: $\epsilon$ is hence
the probability of the agent taking a random, exploratory action (as
opposed to taking the current best, exploitatory action). Each ally
agent is in a constant state of exploitation, as their $\epsilon$
value is 0 for the entirety of the lifetime. Following the choice
of action, the agent carries it out and receives a reward based on
the reward function. The leader agent stores the experience (state,
action, reward, next state) in its replay memory. These stored experiences
are used to periodically update the DQN, enhancing the agent\textquoteright s
decision-making capabilities. For the ally agents, the reward values
are used to determine the role they play in the environment. Following
the learning phase, the DQN agent and the allies are evaluated for
a period of $n_{eval}$ episodes. This evaluation period contains
no learning and is fully exploiting the agents models, due to $\epsilon$
being set to 0, to ensure the choice of action with the maximum estimated
Q-value at every time step. 

The environment parameters used for training and evaluation are shown
in Table I. These values were chosen to create a competitive landscape
that allows for the evaluation of the learning model and strategy.
The environment is set up to be a resource-constrained ecosystem,
with a resource density of 0.1, where the total adversarial agents
present is equal to the total of leader and ally agents. The grid
size was chosen as $100\times100$ to provide a sufficiently large
area for strategy development while still allowing for reasonable
execution times. The parameters that determine the functionality of
both the leader and ally agents are outlined in Table II.

Traditional Q-value updating formula 

\[
Q(s,a)\leftarrow Q(s,a)+\alpha[r+\gamma\max_{a'}Q(s',a')-Q(s,a)]
\]
can be used to make sense of parameters; Q-function is approximated
by the deep neural network, as we are performing DQL. $\epsilon$
changes over time with a decay equation $\epsilon(t)=\epsilon_{0}e^{-\Delta t}$and
is kept above a defined minimal value $\epsilon_{min}.$

\begin{minipage}[t]{0.4\columnwidth}%
\begin{table}[H]
\caption{Environment Parameters}

\centering{}%
\begin{tabular}{|c|c|}
\hline 
Parameter & Value\tabularnewline
\hline 
\hline 
Grid size & $100\times100$\tabularnewline
\hline 
Density $\rho$ & 0.1\tabularnewline
\hline 
Leaders & 1\tabularnewline
\hline 
Allies & 4\tabularnewline
\hline 
Adversaries & 5\tabularnewline
\hline 
\end{tabular}
\end{table}
\end{minipage}\quad{}%
\begin{minipage}[t]{0.45\columnwidth}%
\begin{table}[H]
\caption{Agent Parameters}

\centering{}%
\begin{tabular}{|c|c|c|}
\hline 
Parameter & Leader & Ally\tabularnewline
\hline 
\hline 
Learning Rate $\alpha$ & 0.01 & N/A\tabularnewline
\hline 
Discount factor $\gamma$ & 0.8 & N/A\tabularnewline
\hline 
Initial epsilon $\epsilon_{0}$ & 1 & 0\tabularnewline
\hline 
Decay rate $\Delta$ & 0.995 & 0\tabularnewline
\hline 
Epsilon min $\epsilon_{min}$ & 0.1 & 0\tabularnewline
\hline 
Visible area radius & 10 & 10\tabularnewline
\hline 
\end{tabular}
\end{table}
\end{minipage}

\medskip{}

\subsection{Experiments}

\subsubsection{Model sharing}

Firstly, an analysis is done on the impact on the frequency of model
sharing, by varying the number of episodes per lifetime, while keeping
the total episode count constant. The total episodes used for training
is 400, and the values used for episodes per lifetime are then varied.
The agents are then evaluated against the adversarial agents for 50
episodes and the number of episodes in which the DQN agent and the
ally agents collect a greater number of resources than the adversarial
agents is recorded. This is the performance metric when comparing
the frequencies of model sharing.

Varying the model sharing frequency facilitates understanding the
impact of model sharing on the collective learning curve and strategic
synchronization of the agent network. With more frequent model sharing,
allies quickly receive updates from the learning agent, potentially
leading to a rapid alignment with the leader\textquoteright s current
strategy. This could result in less diversity in role specialization
initially, as all allies quickly converge on the strategies proven
effective by the learning agent. However, it also means that any strategic
adjustments or role shifts identified by the learning agent are promptly
transferred to the allies, allowing for a dynamic and responsive collective
strategy. Conversely, less frequent model sharing allows ally agents
more time to operate independently between updates, potentially fostering
a broader exploration of the environment and the strategic space afforded
by the reward function. This could encourage a greater diversity in
role specialization, as allies might engage in different aspects of
the environment or interact with adversarial agents in varying capacities,
leading to a more heterogeneous set of strategies within the collective. 

\begin{figure}[tb]
\begin{centering}
\textsf{\includegraphics[scale=0.57]{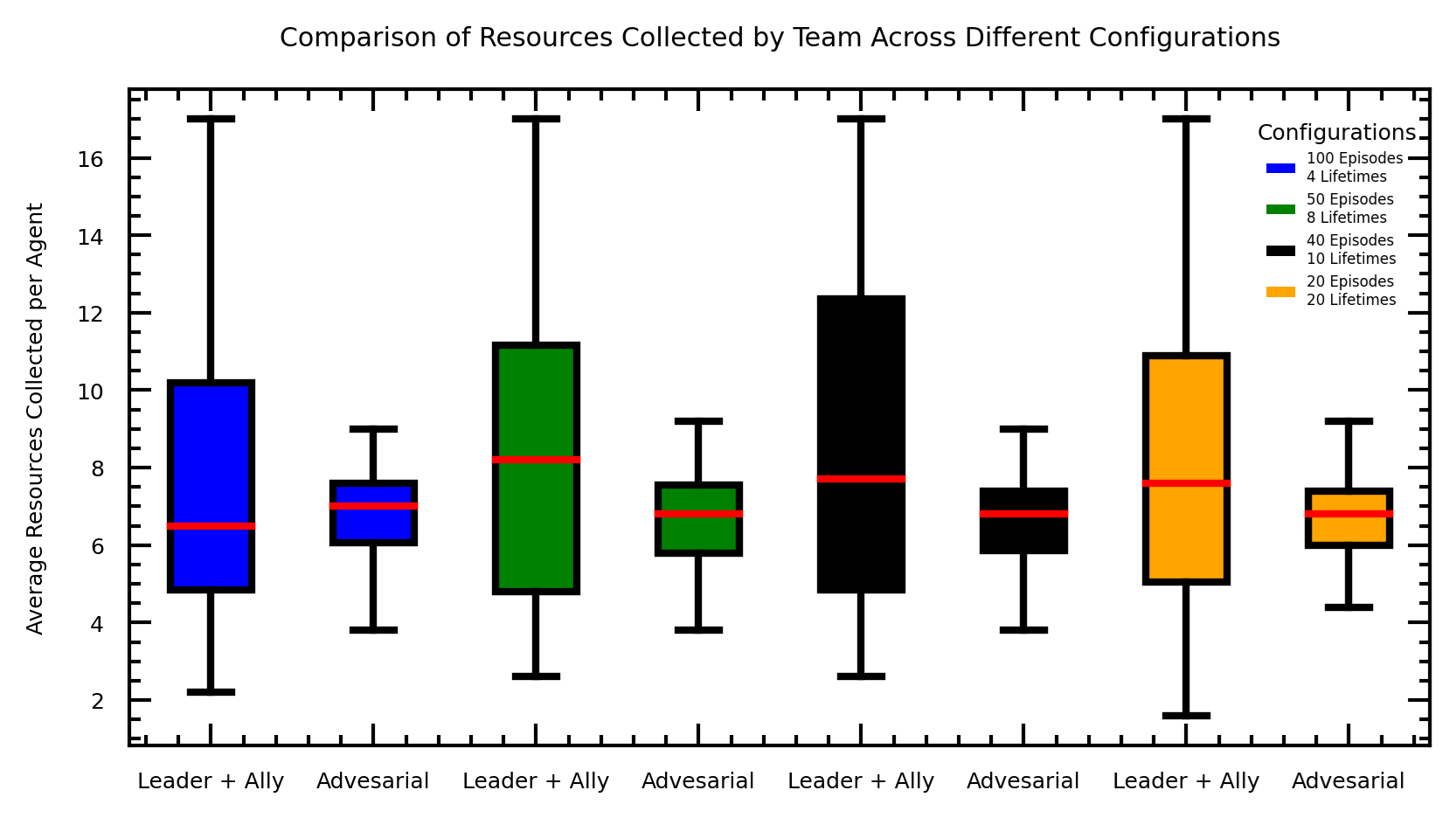}\label{f61}}
\par\end{centering}
\caption{Average and variance of resource collection for each team for each
configuration of episodes and lifetimes during the 50 evaluation episodes.}
\end{figure}

\begin{figure}[htbp]
\begin{centering}
\textsf{\includegraphics[scale=0.5]{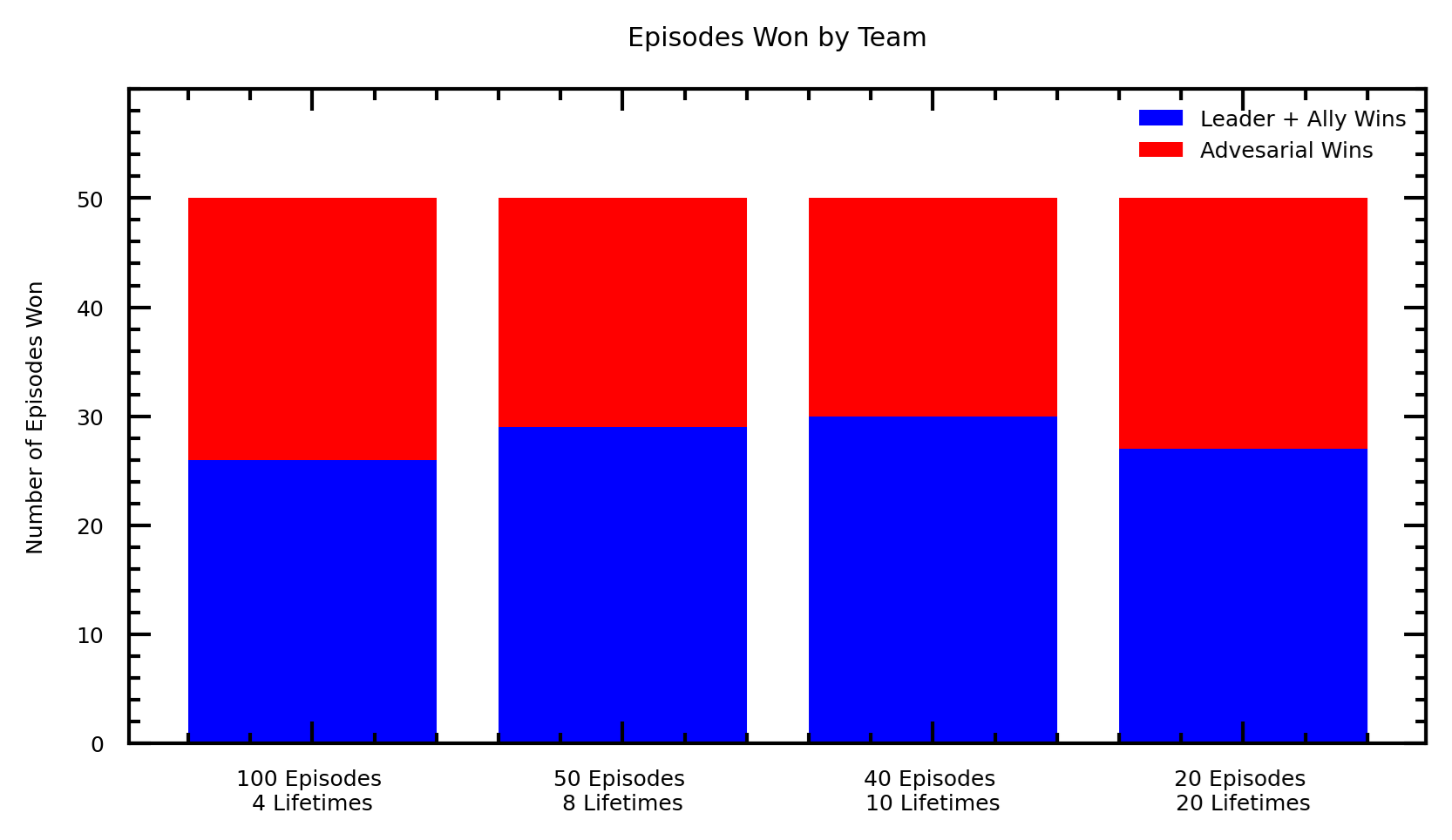}\label{f62}}
\par\end{centering}
\caption{Number of successes by each team across the 50 evaluation episodes
for each configuration of episodes and lifetimes.}
\end{figure}

The results of varying model sharing frequency are shown in Figs.
1 and 2. These results depict the performance benefit that is found
with an increase of model sharing frequency. All cases of model sharing
outperform the adversarial agents in the environment, all winning
more than 50\% of game and averaging more resources collected per
agent. However, the highest model sharing frequency, 20 lifetimes,
with 20 episodes per lifetimes, does not continue the trend of increases
performance with the increase in model sharing frequency. This suggests
existence of an optimal frequency.

The breakdown of the 5 agents rewards, 1 leader and 4 allies, received
throughout training of the configuration of 10 lifetimes, with 40
episodes can be seen in Figure 3. This result shows the differentiation
of roles among the population. Ally 1 receives 78.64\% of its total
reward from $R_{a}$, more than 10\% greater than the next highest.
This indicates that this acts as more of an explorer agent in the
environment, gathering resources away from other allies, compared
to the other agents. In contrast, the leader and other allies exhibit
a more balanced reward distribution from $R_{a}$ and $R_{e}$, suggesting
a mix of disruptor and explorer behaviors. The emergence of a differentiated
role within the population shows the reward system proposed effectively
promotes role differentiation.

\begin{figure}[tb]
\begin{centering}
\textsf{\includegraphics[scale=0.5]{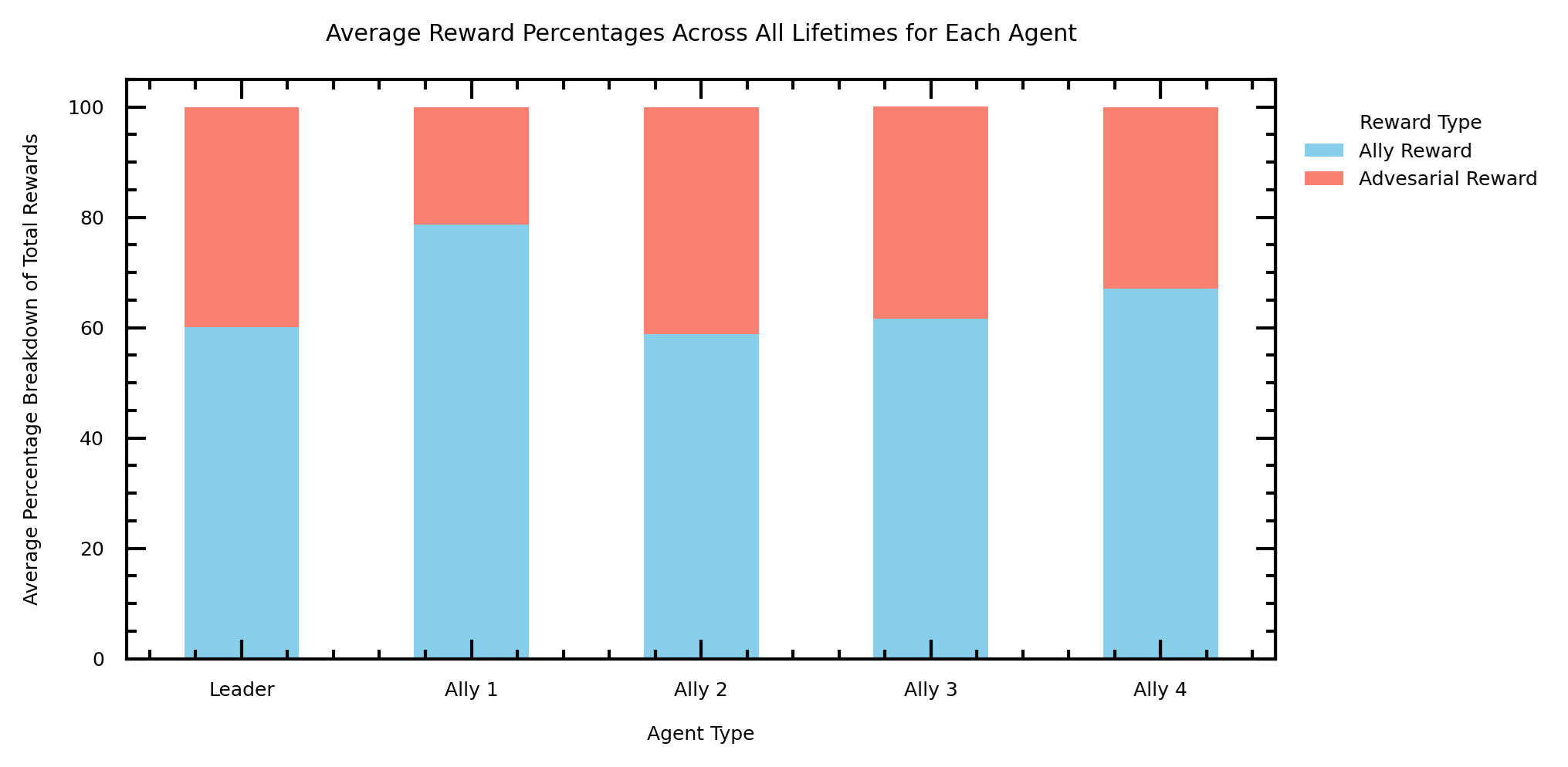}\label{f63}}
\par\end{centering}
\caption{Reward breakdown for leader and allies in the configuration of 10
lifetimes, with 40 episodes per lifetime}
\end{figure}

\subsubsection{Comparison to the state of the art}

An analysis is also completed on the performance of the single learning-agent
approach in comparison to two other traditional approaches. Firstly,
a multi-agent reinforcement learning (MARL) configuration is considered,
wherein each leader agent independently learns and refines its strategy,
absent of model sharing or collaborative learning mechanisms. This
approach represents an autonomous learning environment where each
agent is self-reliant and evolves its policy based on personal interactions
with the environment, unaided by the experiences of ally agents. The
second, a centralised DQN, wherein all agents\textquoteright{} visible
areas are inputted and processed by the network and every agent\textquoteright s
optimal action is outputted. This centralised approach is indicative
of a collective intelligence paradigm where the decision-making process
is unified for the cohort of agents. Both of these strategies are
trained for 400 episodes, just as the single-learning agent strategy
was to ensure consistency in training and a fair comparison. The same
environment parameters for grid size, resource density and adversarial
agents, shown in Table I, are used again in training and evaluating
these comparison models.

\begin{figure}[tb]
\begin{centering}
\textsf{\includegraphics[scale=0.56]{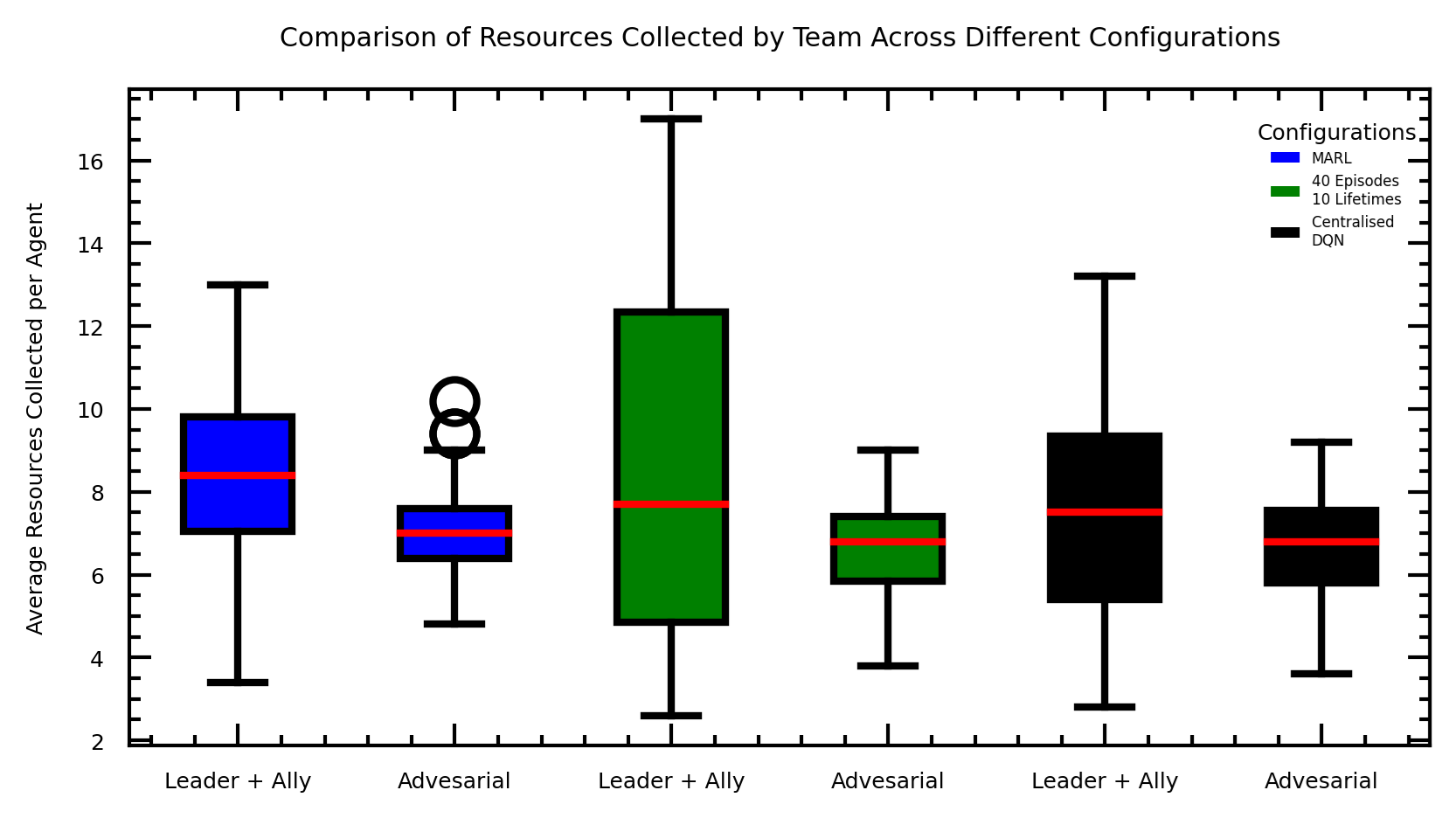}\label{f64}}
\par\end{centering}
\caption{Average and variance of resource collection for each team for each
configuration of episodes and lifetimes during the 50 evaluation episodes.}
\end{figure}

\begin{figure}[tb]
\begin{centering}
\textsf{\includegraphics[scale=0.5]{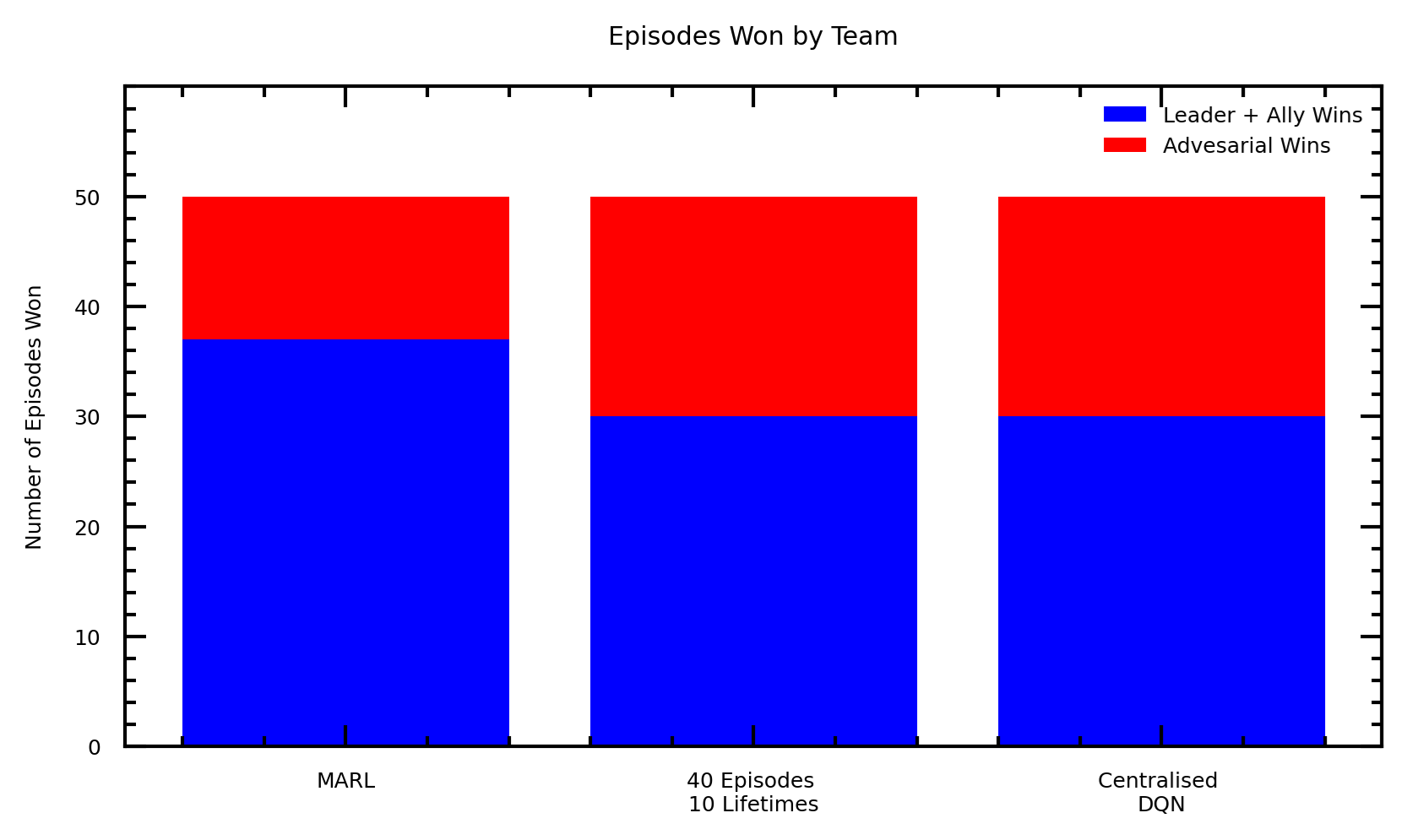}\label{f65}}
\par\end{centering}
\caption{Number of successes by each team across the 50 evaluation episodes
for each RL approach}
\end{figure}

\begin{figure}[tb]
\begin{centering}
\textsf{\includegraphics[scale=0.56]{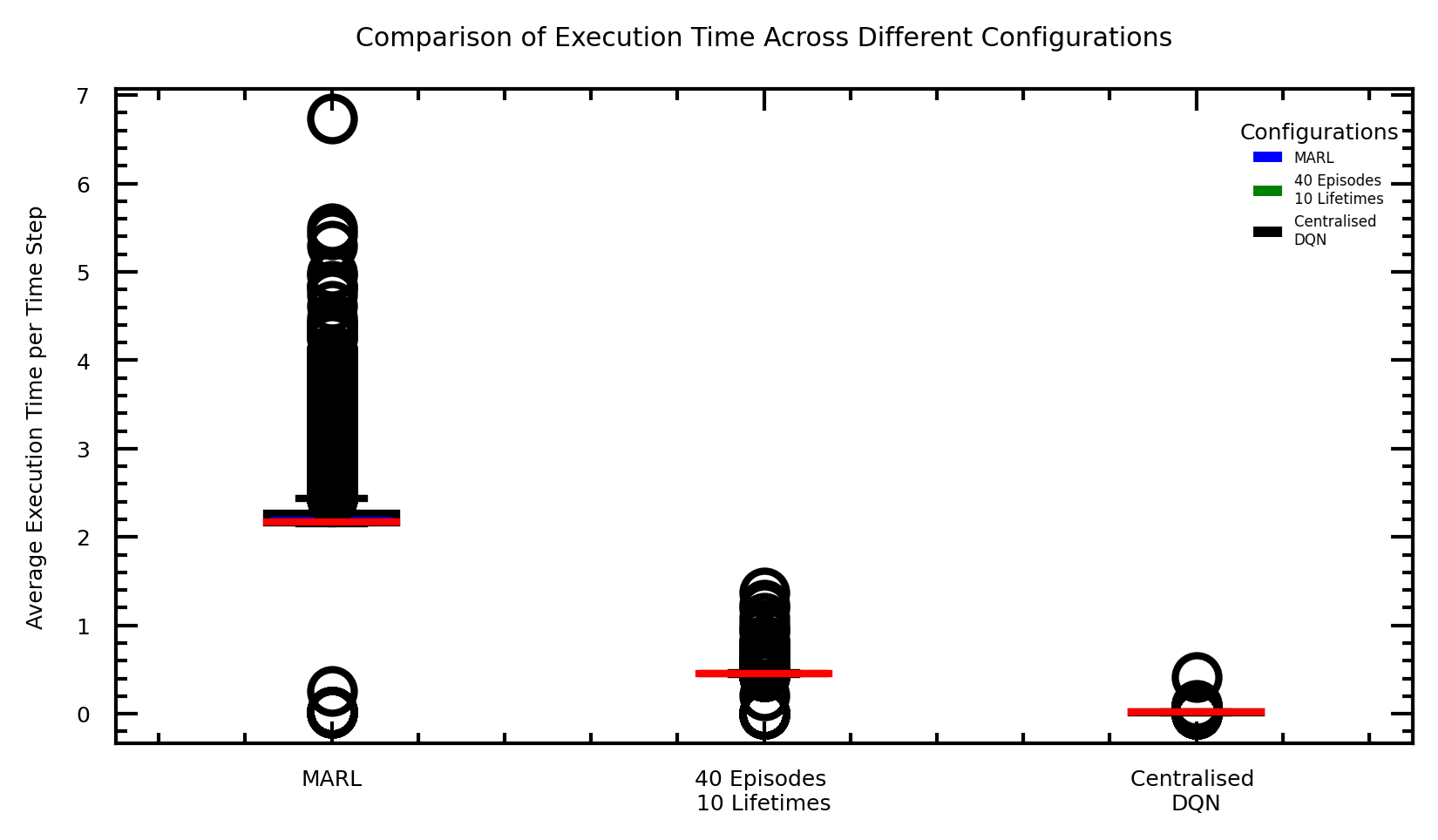}\label{f66}}
\par\end{centering}
\caption{Average and variance of completion time per time step (in seconds)
by each team across the 400 training episodes for each RL approach}
\end{figure}

The results for the two comparison models and the single-agent learning
approach are visualised in Figs 4 and 5. Figure 6 shows the average
and variance execution time of a single time step across the three
approaches.

Comparing the single-agent learning approach with MARL and centralised
DQN model, it can be seen that the MARL approach shows the highest
performance in ally team success but falls short of the average resource
collection performance of the single-agent learning. The variance
observed in the single-agent learning approach\textquoteright s performance,
suggests a broader range of outcomes. It implies that the single-agent
learning model, at its best, outperforms the MARL approach in resource
collection, but it can also fall behind under certain conditions.
This results in a higher average resource collection but a lower success
rate over the adversarial agents for the single-agent learning approach
compared to MARL.

The proposed approach sees a significantly higher performance in average
resource collection compared to the centralised DQN approach, but
performs equally in success rate. This suggests that our approach
is well-suited to scenarios in which higher total resource collection
across all episodes is considered success rather than individual episode
success. Comparing the average execution time per timestep, it can
be seen that the proposed single-agent learning approach significantly
outperforms the MARL approach. The proposed approach also outperforms
MARL in average resource collection, showing its ability to perform
well, with lower training time. However, it does fail to reach the
training speed of the centralised DQN. The results indicate that the
single-agent approach, with its moderate execution times and strategic
dissemination of knowledge, strikes an optimal balance between performance
and execution time. The single-agent model, by design, reduces the
computational burden by having only one agent actively learn and subsequently
share its strategy. This significantly lowers the energy consumption
during the training phase, which is particularly beneficial in scenarios
where energy efficiency is as critical as operational performance. 

In terms of scalability, the centralised DQN can have high computational
overhead associated with managing an increasing number of agents'
state and action information. MARL also sees large, although linear,
increases in computational cost as the number of agents increases
due to the additional agent training required {[}13{]}. This contrasts
to the small increase in computational cost due to model sharing among
additional agents for the single-agent learning approach. This positions
the single-agent learning approach as an energy-conscious solution
that does not compromise on the ability to compete with traditional
MARL strategies in resource collection tasks, offering robust scalability
and flexibility.

\section{Conclusions}

The results from the experiments conducted on the proposed single-agent
learning approach show its performance capabilities can meet the benchmark
set by traditional approaches such as multi-agent reinforcement learning
and a centralised learning model. By centralising the learning to
a single agent and distributing the model to the other agents, the
approach can see these levels of performance with significantly less
computational overhead, with the mean timestep of the proposed solution
requiring 20\% of the time that MARL requires. The results also point
to the presence of an optimal model sharing frequency due to the initial
increase in performance with increase in frequency followed by the
decrease in performance for the higher frequency configurations. Furthermore,
the integration of innovative reward function, allows for the successful
emergence of specialised roles among agents, such as resource explorers
and adversarial disruptors. This role differentiation facilitates
the enhanced collective performance without predetermined roles. These
results highlight the potential of integrating energy aware AI deployments
by providing reduced computational demands that do not compromise
performance compared to traditional methods. Due to the low computational
demand, this approach has the potential for deployment in applications
where efficient resource use is critical, such as embedded systems
in autonomous robotics and IoT networks.

In our future work, we will investigate generalisation options for
the reward function, its sensitivity and and specific conditions under
which roles emerge. Furthermore, we will investigate scalability of
the proposed learning mechanism, and its performance on more complex
environments, scenarios, and realistic considerations of sensor and
actuator limitations in robotic teams while quantifying the computation
and communication cost of learning and coordination.

\end{document}